# Exploring the potential of AI in nurturing learner empathy, pro-social values and environmental stewardship

Kenneth Y T Lim

National Institute of Education, Singapore

Minh Anh Nguyen Duc, Minh Tuan Nguyen Thien

Independent authors

**Abstract:** With Artificial Intelligence (AI) becoming a powerful tool for education (Zawacki-Richter et al., 2019), this chapter describes the concept of combining generative and traditional AI, citizen-science physiological, neuroergonomic wearables and environmental sensors into activities for learners to understand their own well-being and emotional states better with a view to developing empathy and environmental stewardship. Alongside bespoke and affordable wearables (DIY EEG headsets and biometric wristbands), interpretable AI and data science are used for learners to explore how the environment affects them physiologically and mentally in authentic environments. For example, relationships between environmental changes (e.g. poorer air quality) and their well-being (e.g. cognitive functioning) can be discovered. This is particularly crucial, as relevant knowledge can influence the way people treat the environment, as suggested by the disciplines of environmental neuroscience and environmental psychology (Doell et al., 2023). Yet, according to Palme and Salvati, there have been relatively few studies on the relationships between microclimates and human health and emotions (Palme and Salvati, 2021). As anthropogenic environmental pollution is becoming a prevalent problem, our research also aims to leverage on generative AI to introduce hypothetical scenarios of the environment as emotionally strong stimuli of relevance to the learners. This would provoke an emotional response for them to learn about their own physiological and neurological responses (using neuro-physiological data). Ultimately, we hope to establish a bidirectional understanding of how the environment affects humans physiologically and mentally; after which, to gain insights as to how AI can be used to effectively foster empathy, pro-environmental attitudes and stewardship.

## 1. Introduction

Climate change has been one of the most urgent problems to confront in the 21st century. The Sixth Assessment Report (AR6) of the Intergovernmental Panel on Climate Change (IPCC) highlights the unequivocal role of human activities in causing widespread and rapid changes occurring in the Earth's climate systems, some of which may be irreversible [1]. Climate change has been degrading the quality of life for every creature on Earth. Glaciers have shrunk, ice on rivers and lakes is breaking up earlier, plant and animal ranges have shifted and trees are flowering sooner [2]. A microclimate is a small area within the surrounding larger area with a different climate [3]. Any given climatic region therefore comprises many other types of microclimates, which vary in characteristics from the region as a whole. Comparing the human scale to that of the various habitats in which we live, the difference of these scales, compounded by rapid climate change, means that changes in the climates of these habitats may disproportionately affect the conduct of our daily activities. As of now, the detrimental effects of anthropogenic environmental pollution have worsened the physical environment for education, teaching, and learning. For instance, heat waves have been shown to seriously impair students' health and productivity [4]. Palme and Salvati (2021) highlighted that there is relatively inadequate research focusing on the connections between microclimates and human health and emotions [5]. Therefore, it is crucial to delve into how microclimates influence health and productivity to tackle this overlooked issue. In June 2022, the IPCC suggested that rapidly increasing climate change presents a growing risk to mental health and psychosocial well-being, from emotional distress to anxiety, depression, grief, and suicidal behavior. Thus, the investigation of electrodermal activity (EDA)—referring to the continuous variation in the electrical characteristics of the skin, which varies with the moisture level—as a noninvasive method to detect stress and emotional arousal is an area of interest. EDA is linked to the sympathetic nervous system and consists of two components: tonic and phasic, which are represented by skin conductance level (SCL) and skin conductance response (SCR) [6]. For the frequency domain, features relative to EDASymp, TVSymp (spectral powers in specific frequency bands according to Posada-Quintero et al. (2016a; 2016b) [7, 8] and their normalized versions were focused on as they were found to be highly sensitive to orthostatic, cognitive, and physical stress (Posada-Quintero et al., 2020) [9]. It is self-evident that climate change has various effects on individual well-being. Specifically, climate change could potentially alter microclimates to a degree where such changes could adversely affect the physical and mental well-being of those residing in these environments. For instance, a study by Liu et al. in 2019 concluded that "the increasing research interest in thermal comfort and health has heightened the need to figure out how the human body responds, both psychologically and physiologically, to different microclimates" [10]. Hence, delving into EDA data may reveal previously implicit connections regarding how microclimate correlates with our perception of well-being on a detailed scale.

With the rise of Artificial Intelligence (AI), we see the opportunity to use AI in understanding well-being and emotional states better with a view to developing empathy and environmental stewardship. AI is computer algorithms capable of performing difficult tasks that traditionally require human intelligence (such as object identification, drawing relationships and pattern identification) [11]. Using labelled data, in our case, machine learning models were trained to potentially recognise complex relationships between the changes in environment and his own health, stress and emotion, which can be measured by indicators such as EEG and EDA. This understanding can help to raise environmental awareness. In addition to this traditional approach of relationship analysis, we proposed another application with the use of generative AI. Generative AI are deep learning models that can create new contents, such as images, texts, simulations and audio. According to Chui et al. in 2023, Generative AI has become a global phenomenon, with widespread usages in various areas and purposes [12]. There have also been changing perceptions in various fields with the use of Generative AI. With regards to this, we are particularly interested in Generative AI, specifically image generation and content fill's potential in understanding how it can be a stimuli to elicit emotions and change perceptions. Made possible by the measurements and analysis of this was made possible via various indicators like EDA and EEG, our goal is to understand the extent to which Generative AI can effectively foster empathy, pro-environmental attitudes and stewardship.

In our study, two types of experiments were devised. One experiment focused on how the usage of AI could allow us to understand the relationships between the environment, stress and emotions while doing cognitively stressful tasks. The other type of experiment was designed to gain insights into the effectiveness of Generative AI, specifically image generation and content filling as a stimulus that can elicit strong emotional response to environmental degradation. This can unveil insights into Generative AI as an engine for empathy, pro-environment attitudes and stewardship. We see our potential contribution to the notion of human-AI symbiosis, when AI can be a tool to process data and give meaningful insights, at the same time a potential stimulus as a cause for change. This also has implications on improving the living environment for better well-being, as well as instilling pro-environmental attitudes that can contribute to reducing environmental degradation and, even better, improving it.

## 2. Literature review

### 2.1. Urgency of Studying the Effects of Microclimates in the Context of Rapid Climate Change

The phenomenon of rapid urbanization, particularly notable in developing nations, has spurred significant migration flows toward urban centers [13]. As per Statista, global urbanization reached approximately 56% in 2020 [14]. With the pace of urbanization accelerating, alterations to urban environments and climates are inevitable [15]. This widespread urbanization complicates predictions regarding anthropogenic impacts on Earth's climate. At local levels, activities linked to changes in land use, land cover, and urban expansion result in various impacts, including alterations in atmospheric composition, water and energy balances, and ecosystem dynamics [16]. Given the interconnected nature of ecosystems, even minor changes in one component can trigger nonlinear effects elsewhere. For instance, a study by Xiong et al. in 2015 examined the effects of different air temperature shifts on human health and thermal comfort, revealing sensitivities such as perspiration, eye strain, dizziness, accelerated respiration, and increased heart rate as reported symptoms [17].

Amidst global climate change and the exacerbation of urban heat island effects, urban living conditions have deteriorated, significantly affecting human thermal comfort and health [10]. These effects extend beyond psychological impacts, influencing thermal sensation, mood, and concentration, to physiological repercussions such as sunburn, heat stroke, and heat cramps. Liu et al. (ibid.) have also cautioned that "global climate change and intensifying heat islands have reduced human thermal comfort and health in urban outdoor environments". The mentioned effects extend to the context of teaching/learning and classroom environment. For example, heat waves have been shown to seriously impair students' health and productivity [4]. One way to explain this is that the external environmental setting can be an ambient stressor on students' health [18]. Since climate change results in the continuous change in the external environment, e.g., the microclimate inside a classroom, a different degree of stress might be experienced by students, even teachers. With a deeper understanding of the extent to which microclimatic change can affect health, stress, and performance, measures can be taken to maximize productivity

### 2.2. Utility of Electrodermal Activity (EDA) Data in Research

As defined by Critchley and Nagai [19], electrodermal activity (EDA) denotes a measurement of neurally mediated effects on sweat gland permeability, manifested as alterations in skin resistance to a minor electrical current or variations in electrical potential across different skin regions. EDA comprises both tonic and phasic components, represented, respectively, by skin conductance level (SCL) and skin conductance response (SCR). It is closely linked to human stress and emotional responses and can be measured non-invasively [6]. Hence, it is a viable option for reliable and accurate assessment of human stress and emotions in a more comfortable setting outside laboratories. Ward et al. in 2004 [20] have demonstrated that electrodermal activity and heart rate variability are under the influence of the responses of the autonomic nervous system to psychological and emotional activity. They have suggested that any change in EDA during so-called sustained attention to response tests (SARTs) would also be reflected in a change in heart rate variation. As such, their work suggests the possibility of underlying correlations between electrodermal activity and other biometric factors. EDA data have proven effective in many studies that involve emotional and stress assessment. What is more noteworthy is that there have been attempts to study EDA with a multidisciplinary approach. For example, the effects of thermal variance on a person's electrodermal activity following the circadian rhythm were investigated by Kobas et al. [21]. Another exemplification would be the work of Fernandes et al. in linking physical and social environments with mental health using EDA data to assess emotions [22]. As such, the wide adoption of EDA data in investigating various relationships between humans and the environment means that it is a popular and suitable data type to collect. As argued in the preceding paragraphs, changes in climate can affect individuals both physiologically and psychologically. Our review of the literature also suggests correlations between the changing conditions of the climate and changes in recorded EDA. As such, we posit that an understanding of electrodermal activity contributes to a better understanding of mental well-being with respect to changes in microclimate.

## 2.3. Artificial Intelligence (AI) and its applications

According to Sheikh et al. 2023 study on the definition and background of Artificial Intelligence, Artificial Intelligence (AI) is computer algorithms, with the ability to mimic human intelligence in doing complex tasks, such as pattern identification, object identification and language processing. It also comes with a "degree of autonomy" [11]. In the context of our research, AI takes on the approach of "connectionism", following closely to the neurons functioning in the human brain. Trained on labelled data, with predetermined sets of tasks and rules, AI can then take on calculation and resource intensive tasks. The transformative applications of AI could be seen across various fields, such as medicine technological innovation (Wang et al., 2015) and human performance enhancement (Wilson & Daugherty, 2018) [23, 24]. Over the years, from a rather vague definition of AI, both advancements in understanding of human intelligence and artificial intelligence have significantly pushed the boundaries of applications of AI in our lives. This will further reinforce human-AI symbiosis as suggested by Moravec's Paradox, where tasks that are very difficult for humans, requiring immense calculation power are easy for computers, but sensorimotor and perception tasks are difficult for computers [25].

In recent years, rapid advancements in AI have been driven by the usage of machine learning techniques (ML) and innovations in Deep Learning (DL) (Sheikh et al., 2023) [11]. Towards the end of 2022, Generative AI became a worldwide phenomenon with the introduction of "Chat Generative Pre-trained Transformer" or ChatGPT by OpenAI. Generative AI refers to Deep Learning models that are capable of generating multimodal contents such as texts, images, audios and simulations. There are many types of Generative AI, from text-to-image generative models (Qiao et al. 2022), text-to-speech generative models (Chowdhury & Hussan, 2023) to recently, text-to-video generative models (Singer et al., 2022) [26, 27 ,28] . It is also common to find these Generative AI tasks, such as ChatGPT for text-to-text generation, Midjourney for text-to-image and Sora for text-to-image generation. Generative AI has had a tremendous effect on various sectors. In education, it fundamentally changed how lessons could be delivered and how knowledge could be obtained. In a study by Moorhouse & Kohnke in 2024, the result showed that Generative AI tools would substantially alter the Initial Language Teacher Education curriculum in terms of incorporating Generative AI into learning outcomes, and AI literacy [29]. This means that there will be a significant change in perceptions of how education is delivered in conjunction with generative AI. In another study by Firat in 2023, Generative AI could encourage a shift in mindset to autodidactic learning, as tools like ChatGPT can provide feedback and answers to students, void of criticisms or biases [30]. In another 2024 study of consumerism by Mogaji & Jain, Generative AI could change consumer behaviour in various aspects, such as giving consumers recommendations, changing purchasing behaviours and shaping global trends [31]. In both of these cases, generative AI has shown the potential to change perceptions of stakeholders.

According to a study by Lin et al. in 2024, Generative AI, specifically image generation, is able to affect or effectively elicit emotions via emotional image editing [32]. There have been rigorous studies in various global and local factors that can elicit certain emotions, such as different tones of backgrounds, and object associated emotions, like graveyards associated with sadness and balloons associated with amusement [33]. To elaborate further, Nico (2009) found that certain objects tend to have a relationship with corresponding emotions [34]. In our study, we focused on representing sentiments via two dimensional (2-D) valence-arousal coordinate space (Nicolaou et el., 2011)[35]. The emotional valence dimension helps classify pleasantness of an emotion ranging from positive or pleasant to negative or unpleasant [36]. The emotional arousal dimension helps classify the degree of intensity of emotions [37]. This has implications for generating impactful images that convey the bleakness of environmental degradation and climate change. Furthermore, in a 2020 study on emotional responses and plastic waste, Septianto & Lee concluded that images that depict plastic wastes could be tailored to target various emotional evocations, such as disgust and sadness [38]. They also demonstrated how this understanding of discrete emotional responses could reduce consumers' impact on the environment as well as propose policies for social marketers and policy makers. Thus, by leveraging the power of Generative AI in generating impactful environmental images, those that convey messages of degradation and pollution through symbolic objects like trash, there can be powerful emotions experienced by viewers. These emotions can catalyse transformative shifts in mindsets and perceptions that could fundamentally foster pro-environment attitudes and stewardship. According to Haller, empathy for the environment would encourage more willingness to take action to mitigate climate change. Not only are there implications for policy makers and marketers, individuals themselves can also be directly influenced to change for the better [39].

## 3. Methodology

### 3.1. Collecting Electrodermal Activity (EDA)

Data For the do-it-yourself (DIY) physiological wristband used in this study, EDA sensors are built based on the design described in Zangróniz et al. (2017) [40] with an input voltage of 3.3 V and a 10 Hz sampling rate. LM324 operational amplifier with low noise of 35 nV/rtHz was used alongside Dry Ag/AgCl Finger Electrodes. TP4056 battery charger circuit and 1200 mAh 3.7 V lithium battery provided approximately 10 h battery life for the unit.

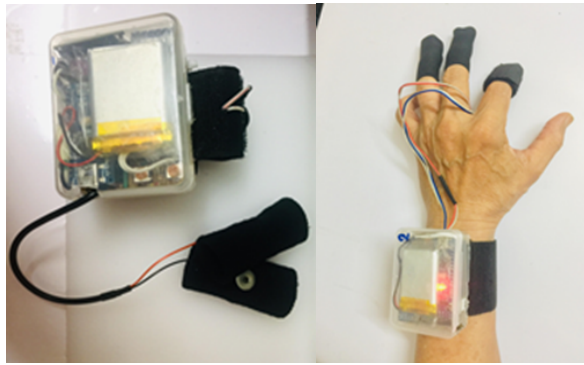

**Figure 1a. and b.** Assembled EDA wristband.

### 3.2. Collecting Environmental Data

To gather microclimate data, a compact portable device was constructed to assess the ambient environmental conditions, including noise level, light intensity via infrared radiation, dust concentration, carbon dioxide concentration, temperature, relative humidity, air pressure, and wind speed. Sampling rate of the unit is roughly 1 Hz. The schematics for the device are depicted in Figure 2a, and the assembled device is shown in Figure 2b.

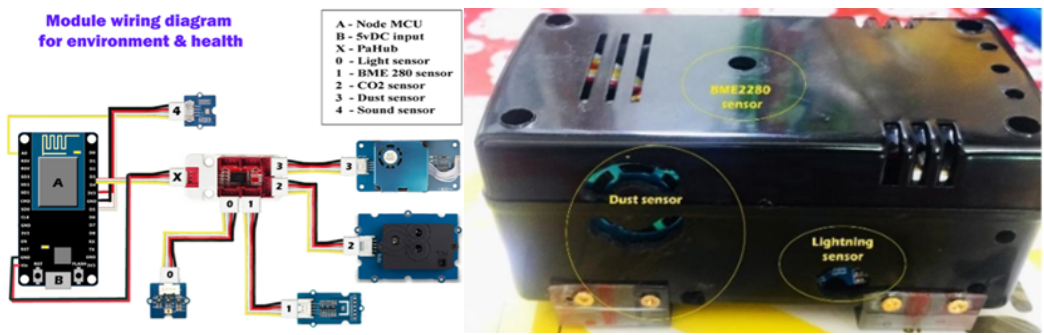

**Figure 2a and 2b.** Schematics for environmental sensors device and assembled device.

### 3.3. Data collection - self-reported measures

Data collection through self-reported measures was carried out through an assessment of general stress, and an assessment of mental stress. A baseline score was determined before any tasks were assigned and under prevailing ambient microclimatic conditions. Participants were requested to evaluate the task difficulty on a scale ranging from 1 to 10. Similarly, they were prompted to rate their stress level using a scale from 1 to 7. The range of the scale was designed to sync with subsequent types of questionnaires for mental and workload assessment. For the purposes of the study reported in this paper, a self-reported score of task difficulty greater than 6, and a level of stress greater than 4 was interpreted as 'high stress'; conversely a score of task difficulty and a level of stress less than 3 was interpreted as 'low stress'.

As for the assessment of mental stress, we chose to focus on what is generally termed acute stress, that is, short-term, event-triggered exposures to threatening or challenging stimuli that elicit a psychological and/or physiological stress response, such as delivering a public speech. To assess this, we administered the instrument known as the Self-Assessment Manikin (SAM) to participants. SAM is a "non-verbal pictorial assessment technique that directly measures the pleasure, arousal, and dominance associated with a person's affective reaction to a wide variety of stimuli" [41]. SAM was chosen for its effectiveness in directly assessing the pleasure, arousal, and dominance associated in response to the event, which reduces resources needed to measure other types of variables for higher resolution. Furthermore, SAM can be used universally as language barriers can be overcome when emotions are represented by pictures.

For stress in this study, we focused on: Acute Stress - Short-term, event-based exposures to threatening or challenging stimuli that evoke a psychological and/or physiological stress response, such as giving a public speech. the Acute Stress Appraisals questionnaire was used. The Acute Stress Appraisal emphasises the multifaceted nature of demand and resource appraisals to be used in laboratory stress paradigms. Demands were defined to be made up of perceived uncertainty, required effort, and how demanding the task seems, among other factors, whereas resources comprise perceived knowledge and abilities, controllability, social support, and expectations [42]. There are two parts to this questionnaire: a pre-task appraisal and a post-task appraisal

### 3.4. Experiment to investigate how the environment affects physiological, mental health and productivity

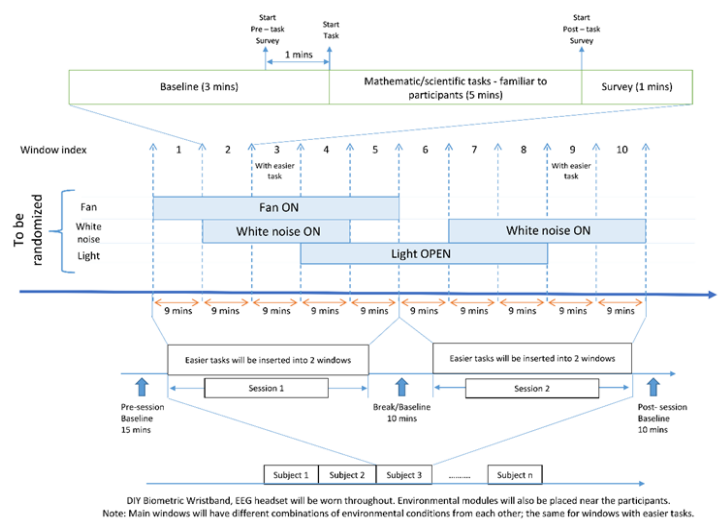

**Figure 3.** Procedure of investigating the effects of environment on stress and emotions

The two-hour period was divided into eight 15 min intervals / windows. Each interval featured varied, randomized combinations of microclimatic factors, as illustrated in Figure 3. Each interval consisted of a 3-minute baseline where participants rest, a 5-minute task and a survey which featured SAM and Acute Stress Appraisal. Participants wore a DIY wristband on the wrist of their non-dominant arm throughout the experiment while engaging in challenging mathematical tasks aimed at sustaining elevated stress levels. Few windows had easier tasks to serve as baseline and low-stress periods. The participants should be junior college students with similar levels of mathematical competency. There are also baseline periods/breaks before, during, and after the experiment.

The collected data underwent processing in Python, with outlier detection conducted using the z-score method to eliminate outlier data. Initially, the collected EDA data were normalized and filtered using a low-pass filter (1.5 Hz, Butterworth, 32nd order) to eliminate unwanted artifacts [43]. Subsequently, the EDA data were decomposed into tonic and phasic components utilizing the convex optimization (cvxEDA) method developed by Greco et al. in 2016 [44]. The number of skin conductance responses per minute (NSSCR) was also derived, representing the quantity of rapid transient events within the phasic component [45]. According to the literature, NSSCR refers to skin conductance responses occurring without a specific eliciting stimulus [46], and NSSCR levels increase alongside ratings of emotional arousal [47]. As per Wichary et al. (2016), heightened arousal and negative valence characterize emotional stress [48]. Thus, this metric serves as a potentially reliable indicator of stress. To conduct frequency-domain analysis, the EDA data were downsampled to 2 Hz. Subsequently, the signals underwent high-pass filtering (0.01 Hz, Butterworth, 8th order) to eliminate any underlying trend. Using variable frequency complex demodulation, TVSymp is calculated as the mean of time-varying spectral amplitudes in the 0.08–0.24 Hz band. In the frequency domain, a Blackman window (length of 128 points) was applied to each segment (0.5-s overlap with each other), and the fast Fourier transform was calculated for each windowed segment. EDASymp(n), as a tool for sympathetic tone assessment, was computed as the normalized power within the frequency band of interest (0.045 to 0.25 Hz) [45]. Random forest regression (RF) models were trained. Random forest regression models performed the best. Thus, it was chosen as the tool of analysis. Random forest regression models are then trained on environmental data and EDA features, with the former as input and the latter as output, with a train–test split ratio of 7:3 to discover the nonlinear relationships between environmental factors and EDA features. The outcomes of the random forest regression models were analyzed using Shapley values and Shapley summary plots to uncover intricate connections between input and output variables.

### 3.5. Experiment to investigate the potentials of Generative AI as a stimulus in fostering environmental empathy and pro-environment attitudes

For the duration of this experiment, participants wore EDA physiological wristbands on the wrist of their non-dominant arm. This was to analyse their emotions, using the aforementioned indicators of EDA in the frequency and time-frequency domains.

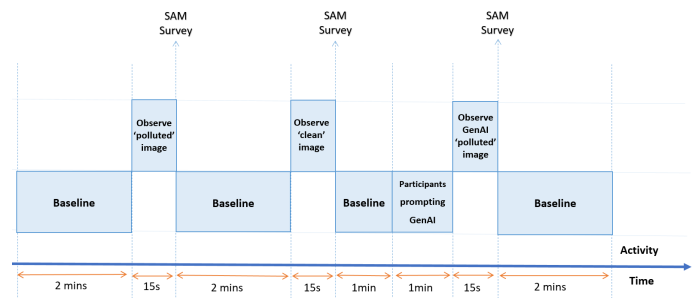

**Figure 4.** Procedure of investigating the feasibility of Generative AI on being a stimulus of change

Figure 4 details the procedure of investigating the feasibility of Generative AI on being a stimulus of change. The experiment began with participants entering a baseline period, during which they were instructed to rest and relax without engaging in any activity for 2 minutes. Following this, participants were shown an image of a polluted environment. They were asked to take a general glance at the image and then focus on any details that caught their attention. Immediately afterward, for 1 minute, participants completed a Self-Assessment Manikin (SAM) survey to assess their emotional response. After this, participants returned to a resting state for another 2 minutes. Next, they observed an image of a pristine environment for 15 seconds (see Figure 5 as an example). They then filled out another SAM survey, which again took 1 minute. After 1 minute of resting, participants moved on to the Generative Fill prompt task. In this task, they were asked to select areas within an image that they wished to generative fill (i.e., generate filled content) and began typing out a prompt for this task on Adobe FireFly. The original image to be altered was one that depicted pristine environments. Participants spent 5 minutes and 30 seconds refining their prompts and selecting the option they were most satisfied with. Afterwards, participants completed a third SAM survey to assess how they mentally feel with the degraded environment, which took 1 minute, and then concluded the experiment with a final 2-minute rest period.

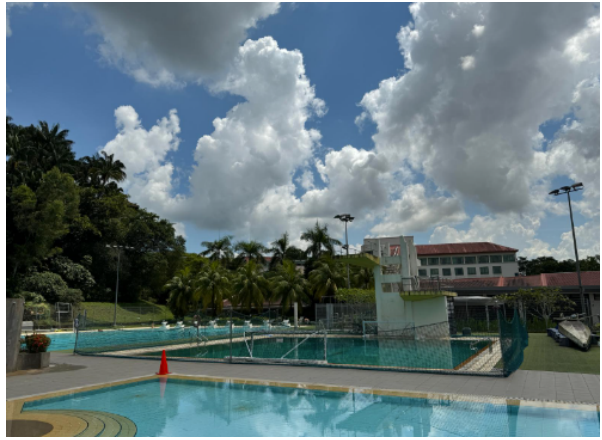

**Figure 5.** Picture of pristine school campus before Generative Fill task

Data processing was the same methodology as the experiment to investigate how the environment affects physiological, mental health and productivity. First, EDA data was preprocessed, then EDA features were extracted. Preliminary statistical analysis on EDA extracted features were performed, such as the Kruskal-Wallis test. Random forest classifiers were trained to detect the emotional states of participants during each observation of pictures, via emotional indicators. Finally, this result was compared against that of the Self-Assessment Manikin data collected.

## 4. Results

For the first experiment to investigate how the environment affects physiological, mental health and productivity, more than 30,000 environmental data points and 300,000 EDA data points were collected from 5 participants (4 males and 1 female). For the second experiment to investigate the potentials of Generative AI as a stimulus in fostering environmental empathy and pro-environment attitudes, data were collected from 4 male participants.

### 4.1. Predicting EDA features using random forest regressor with environmental variables as inputs for high stress windows

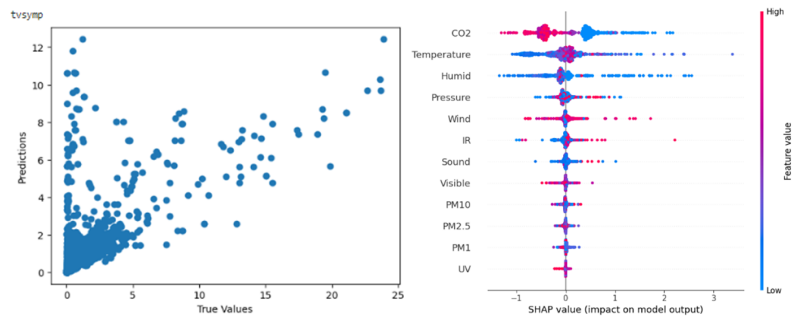

**Figures 6a and 6b.** $R^2$ Score of 0.425 and Shapley Summary Plot using environmental data as input to predict TVSymp

From Figures 6a and 6b, Shapley Summary Plot suggests that TVSymp is most significantly affected by the variable 'CO2 concentration', followed by the variables 'Temperature' and 'Humidity'. For the variable 'CO2 Concentration', higher value of CO2 concentration tends to decrease TVSymp value, which results in lower Stress level and emotional valence. For the variable 'Temperature', higher temperature can either increase or decrease the value of TVSymp. Therefore, it is inconclusive to which extent a change in temperature may affect stress. Finally, for the variable 'Humidity', lower humidity can either increase or decrease the value of TVSymp.

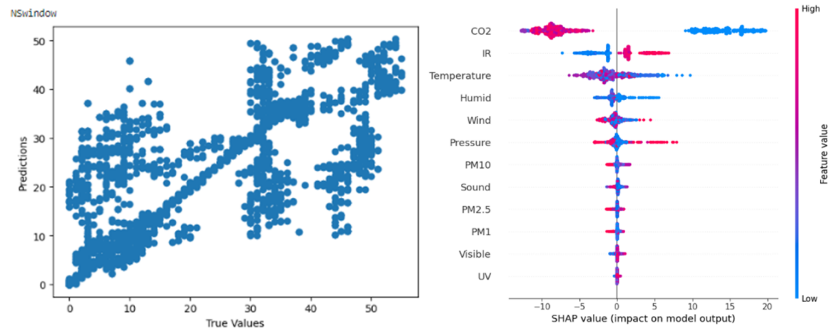

**Figures 7a and 7b.** $R^2$ Score of 0.622 and Shapley Summary Plot using environmental data as input to predict NSSCR

From Figures 7a and 7b, Shapley Summary Plot suggests that the NSSCR is most significantly affected by the variable 'Carbon dioxide concentration', followed by the variables 'Infrared radiation' and 'temperature'. For the variable 'Carbon dioxide concentration', higher carbon dioxide concentration tends to decrease NSSCR value, which in turn correlates with lower Stress level and emotional arousal of participants. For the variable 'Infrared radiation', it can be seen that higher level of infrared radiation (intensity) leads to higher NSSCR value and vice versa. Thus, higher Infrared radiation can lead to higher levels of stress and emotional arousal.

### 4.2. Results of Generative AI as a stimulus to elicit emotional response

#### 4.2.1. Random Forest Classifier to predict emotional arousal, valence and control (dominance)

**Table 1.** EDA input features and its importance in Random Forest Classifiers

|  | Valence | Arousal | Dominance |
|---|---|---|---|
| TVSymp **(n.u.)** feature importance | 0.282 | 0.256 | 0.280 |
| NSSCR **(count/min)** feature importance | 0.397 | 0.260 | 0.397 |
| EDASymp **(µS$^2$)** feature importance | 0.321 | 0.484 | 0.323 |
| **Model Accuracy** | 0.924 | 0.951 | 0.909 |

**Table 2.** Spearman's rho coefficient between EDA features and SAM domains

|  | **TVSymp (n.u.)** | **EDASymp (µS$^2$)** | **NSSCR (count/min)** |
|---|---|---|---|
| **Valence** | - 0.123 | - 0.014 | - 0.198 |
| **Arousal** | 0.332 | 0.434 | 0.577 |
| **Dominance** | 0.097 | 0.326 | 0.192 |

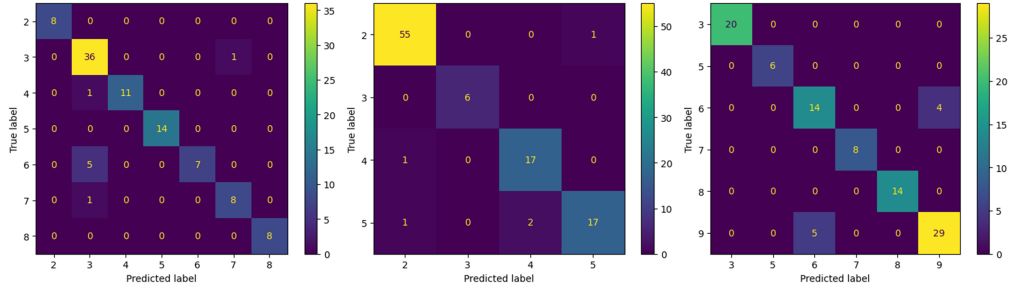

**Fig 8.** Confusion matrices of Random Forest Classifiers (n_estimators=2000) on emotional arousal (left, accuracy = 0.92), valence (middle, accuracy = 0.95) and dominance (right, accuracy = 0.91)

In Fig 8, the higher the number on the True label and Predicted label axes, the higher emotional arousal / valence / dominance. The output of this Random Forest Classifier was the prediction of SAM domains on the Likert scale, of which the labelled data came from the first line of Self-Assessment Manikin. In our experiment, the Likert scale for emotional arousal in the Self-Assessment Manikin (SAM) ranged from 1 to 9. From Table 1, both NSSCR and EDASymp have high feature importance in Random Forest Classifier in predicting discrete emotional arousal. The high accuracy of this classifier means that EDA features could predict emotional arousal accurately. Both EDASymp had highest feature importance in Random Forest Classifier in predicting discrete emotional valence. The high accuracy of this classifier means that EDA features could predict emotional valence accurately. NSSCR had the highest feature importance in Random Forest Classifier in predicting discrete control (dominance). The high accuracy of this classifier means that EDA features could predict control (dominance) accurately.

#### 4.2.2. Statistical analysis to investigate the use of Generative AI to elicit emotions

**Table 4:** EDA features statistics during different events

| Event | TVsymp (Mean ± STD) | EDASymp (Mean ± STD) | NSSCR (Mean ± STD) |
|---|---|---|---|
| **Event 1:** During Baseline | 0.7762 ± 0.7485 | 0.6117 ± 0.1237 | 13.3628 ± 14.2879 |
| **Event 2:** During Generative AI Prompting task | 0.9200 ± 0.8309 | 0.6880 ± 0.1073 | 18.0409 ± 12.3995 |
| **Event 3:** Viewing image of Pristine Environment | 0.8080 ± 0.7557 | 0.6178 ± 0.1180 | 14.8548 ± 13.8581 |
| **Event 4:** Viewing image of Polluted Environment | 1.0719 ± 0.8577 | 0.6462 ± 0.0927 | 19.3034 ± 8.8849 |
| **Event 5:** Viewing generative-filled polluted images | 2.2723 ± 2.6380 | 0.6544 ± 0.1027 | 21.3833 ± 8.0218 |

**Table 5:** p-value of Kruskal Test on EDA features

| Comparison | TVSymp | EDASymp | NSSCR |
|---|---|---|---|
| Rest vs. Affective Stimuli | 1.08E-23 | 1.26E-06 | 1.54E-16 |
| Real Image vs. AI Generated Affective Stimuli | 3.54E-12 | 0.052 | 0.0055 |
| Calming (Pristine) vs. Distressing (Polluted) Stimuli | 4.76E-21 | 1.85E-05 | 3.51E-10 |

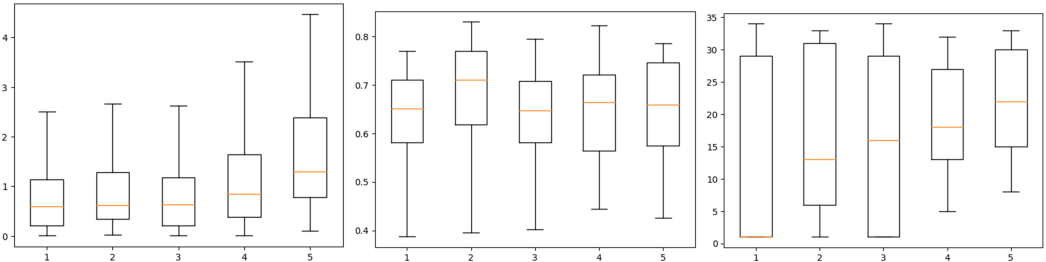

**Figure 9.** EDA features statistical analysis (y-axis: EDA feature, x-axis: EventID)
(In order: left - TVSymp, middle - EDASymp, right - NSSCR)

As seen in Tables 4 and 5, and in Figure 9, TVSymp does not vary much from baseline when participants viewed an image of Pristine Environment and thought of prompts for Generative fill in the image. In contrast, it can be seen that when viewing an original image of Polluted Environment, TVSymp of participants increased by around 0.3. This effect was amplified when participants were viewing polluted images Generatively filled by Firefly from the original pristine environment image. This is statistically significant, as seen in p-value of Kruskal-Wallis Test (3.54E-12). The value of TVSymp during this event (2.2723 ± 2.6380) was more than twice that when the participants viewed only an original, real life captured image of Polluted Environment. This suggests that using Generative AI to create images conveying environmental degradation using a real life image can significantly lead to higher emotional arousal and more negative emotional valence. EDASymp does not vary much from baseline. The value of EDASymp during the Generative AI Prompting task (0.6880 ± 0.1073) and in general, during the periods of viewing images of Polluted environments were the highest. Since there is an increase in EDASymp from baseline, this suggests that Generative AI stimulus has the potential to elicit higher emotional arousal. NSSCR during the event of viewing an image of a pristine environment does not vary much from baseline. In contrast, it can be seen that when viewing an original image of Polluted Environment, during Generative AI Prompting task and viewing images of polluted environment generated with Firefly, NSSCR increased sharply. The value of NSSCR during the event of viewing images of polluted environments generated with Firefly was the highest. Again, this indicates that using Generative AI to create images conveying environmental degradation using a real life image can significantly lead to higher emotional arousal.

## 5. Discussion

### 5.1. Discussion on first experiment to investigate how the environment affects physiological, mental health and productivity

Overall, temperature tends to increase cognitive stress but makes emotional arousal and emotional valence more negative (less positive). Furthermore, it was found that air quality also has a great impact on participants' cognitive stress.

This has demonstrated a universal approach of using the Shapley summary plot to interpret both the direction and magnitude of the effects of microclimate on EDA features. To elaborate, the study has shown that based on the magnitude of Shapley values, the significance and extent to which microclimate affects cognitive stress and emotional states can be inferred. With a known effect of increasing or decreasing the value corresponding to the EDA feature, literature can be reviewed to understand how cognitive stress and emotional state change accordingly. It has solidified the traditional use of AI in understanding stress and emotions. These are the AI tools that will be crucial in raising awareness of the tangible effects of climate change, thus increasing the willingness for people to take action to mitigate climate change and unsustainable practices.

It is acknowledged that the approach in this report has some limitations. There might be a degree of uncertainty in data measurements made by self-built devices and a limited number of participants. Besides improving on these limitations, another key consideration for the future is to consider the significance of circadian rhythm, as it can also affect EDA data. A more expansive multimodal approach could also be devised, such as incorporating data from EEG and facial action units.

### 5.2. Discussion on second experiment to investigate the potentials of Generative AI as a stimulus in fostering environmental empathy and pro-environment attitudes

As suggested by the results of the experiments, the act of viewing images that convey environmental degradation mostly leads to higher emotional arousal and emotional valence. What was even more interesting was the fact that by involving themselves in the act of using Generative AI, specifically Generative fill-in altering images to convey environmental degradation, participants experienced more intense and negative emotions. These emotions could potentially be identified along the emotion spectrum from sadness to disgust.

In this experiment, we acknowledge the limited number of participants. That being said, our study further propounded that Generative AI could be an effective tool in eliciting negative emotions in humans. In combination with previous literature on the psychological effects of these emotions as an engine for change in perceptions and behaviours, Generative AI will play a crucial role in fostering environmental empathy and pro-environmental attitudes. In the future, more experiments will be carried out and other areas of interest can be better studied, such as the extent to which AI will motivate students to learn, or workers to perform their job, whether it is creative or office jobs.

## 6. Conclusions

The research has demonstrated the success of utilising AI and multidisciplinary approach to understanding the link between microclimate and human health, stress, and emotions. This highlighted the feasibility of large scale application of this method to gather larger amounts of data for analysis. This can also be customised to each person, as their emotional and stress responses to changes in the environment can vary. This is backed up by Machine Learning, as the more amount of data there is to train the models, the more accurate it gets in prediction.

Not only has AI proven to be extremely valuable in understanding the relationships between the environment, stress and emotions, but also a catalyst for change. This was demonstrated in Generative AI, more specifically, Generative Fill's ability to elicit certain negative motions that were scientifically proven to psychologically trigger changes in perceptions and behaviours on the environment. Hence, there is tremendous possibility in harnessing the transformative power of AI in evoking emotions to nurture empathy, pro-social values and environmental stewardship.

If utilised correctly on a large scale, AI is the key to a favourable societal shift in mindsets towards sustainability and conservation on a macro level. This has implications for policy makers, marketers and AI developers. Using a closed-loop approach - AI feeding data to users and users providing data to AI models, stakeholders can tailor the training and output of AI to align to these understandings of usage of AI. In combination with other usages of AI in the same topic of the environment such as setting trends and consumer's behaviours, we can hope that in the near future, there will be higher awareness of the environment and pro-environmental practices. This is one of many benefits that human-AI symbiosis will bring in the future.